\documentclass[showkeys,amsmath,amssymb,superscriptaddress,nofootinbib, preprint]{revtex4-1}

\usepackage{booktabs}
\usepackage[bookmarks=true,bookmarksnumbered=true,bookmarkstype=toc]{hyperref} 

\usepackage[pdftex]{graphicx,color}
\usepackage{epstopdf}
\usepackage[bookmarks=true,bookmarksnumbered=true,bookmarkstype=toc]{hyperref} 
\usepackage{units}
\usepackage{slashed}
\usepackage{xcolor}
\usepackage{ulem}

\usepackage{bm}

\preprint{ver. \today}

\begin{document}


\title{Directional Direct Detection of MeV Scale Boosted Dark Matter in Two Component Dark Matter Scenario via Dark Photon Interaction}


\author{Keiko~I.~Nagao}
\email{nagao@ous.ac.jp}
\affiliation{Department of Physics, Okayama University of Science, Okayama 700-0005, Japan}

\author{Tatsuhiro~Naka}
\email{tatsuhiro.naka@sci.toho-u.ac.jp}
\affiliation{Department of Physics, Toho University, Chiba 274-8510, Japan}
\affiliation{Kobayashi-Maskawa Institute for the Origin of Particles and the Universe, Nagoya University, Nagoya 464-8602, Japan}

\author{Takaaki~Nomura}
\email{nomura@scu.edu.cn}
\affiliation{College of Physics, Sichuan University, Chengdu 610065, China}

\begin{abstract}
This study explores a two-component dark matter model in which one component, heavier dark matter, annihilates into a lighter dark matter. The lighter dark matter is expected to generate detectable signals in detectors due to its enhanced momentum, enabling direct detection even for MeV-scale dark matter. 
We investigate the effectiveness of directional direct detections, especially the nuclear emulsion detector NEWSdm, in verifying these boosted dark matter particles through nuclear recoil. 
In particular, we focus on light nuclei, such as protons and carbon, as suitable targets for this detection method due to their high sensitivity to MeV-scale dark matter. 
By modeling the interactions mediated by a dark photon in a hidden U(1)$_D$ gauge symmetry framework, we calculate the expected dark matter flux and scattering rates for various detector configurations.
Our results show that nuclear emulsions have the potential to yield distinct, direction-sensitive dark matter signals from the Galactic center, providing a new way to probe low-mass dark matter parameter spaces that evade conventional detection methods.
\end{abstract}

\maketitle

\clearpage 
%
\section{Introduction}
\label{sec:intro}
The nature of dark matter (DM) is one of the biggest open questions in physics.
Astrophysical and cosmological observations proved its existence through gravitational effects and have revealed the amount of relic density $\Omega_{\rm DM} h^2 \simeq 0.12$ \cite{Planck:2018vyg}.
Theoretically, there are various candidates of DM from very light particles 
including wave-like DM 
to heavy primordial black holes. 
One of the attractive candidates is a particle DM interacting with the standard model (SM) particle weakly since its relic density can be naturally explained by a thermal process. 
Such a particle DM is extensively searched for by direct detection, indirect detection, and collider experiments, especially for DM mass around the electroweak scale.
However, we have not obtained clear evidence of DM particles in these experiments, and an extension of the search area for DM mass and interaction is desired.
In addition, it is worth considering the non-standard signal of a particle DM, which would 
have been missed in previous observations. 

The boosted DM scenario is promising for exploring parameter spaces that are difficult to explore with conventional direct DM searches.
In particular, since the usual direct search experiments target what makes up the current Galactic halo, their velocities are limited by the Galaxy's escape velocity. Therefore, searches for light DM on the MeV scale are particularly difficult because its kinetic energy is below the energy threshold of the detector. Actually, the thermal relic scenario of the DM represented by the Weakly Interacting Massive Particles (WIMP) model, mass scale of MeV/$c^{2}$ is still allowed \cite{DM_CMB:2013,DM_BBN:2019}. While such a lighter mass region can be searched using the electron recoil and the Migdal effect, independent verification of direct interactions with baryons is still needed. 
If the DM-accelerating process exists, the light DM would be detectable in existing detectors, thus making it possible to verify parameter spaces and specific particle physics models that are difficult to search for in the conventional direct search scenario.

For example, various phenomenological studies and experiments have already been carried out in the scenario of DM accelerated by cosmic rays (\textit{e.g.}, \cite{CRDM:2019,PandaX:2022,Nagao:2022azp,SK:2023}). In this scenario, accelerated DM has the broad kinetic energy spectrum of DM up to the GeV scale; an existing neutrino detector with a large scale can search that~\cite{SK:2023}.
Constraints for models with scalar and vector mediators are discussed in ~\cite{Elor:2021swj}.

The multi-component DM scenario, which is the main focus of this study, allows for multiple components in the dark sector.
In such a scenario, boosted DM can be realized when we have heavier DM component that annihilates into lighter one(s)~\cite{Agashe:2014yua,Elor:2015tva,Aoki:2018gjf,Li:2023fzv,Basu:2023wgo}. 
It is assumed that the signal can be detected due to the large momentum of secondary dark particles from the annihilation of the heavier component in the Galactic halo, which dominates relic density.
This scenario is expected to explain the current situation of direct DM searches, which place strong constraints on the interaction cross section with SM particles, particularly those coupling with baryons, for the main DM component in the Galactic halo.
As a concrete model, we adopt the hidden $U(1)_D$ model in ref.~\cite{Li:2023fzv} that contains two components of DM that are Dirac fermions stabilized by the hidden $U(1)_D$ charge assignment. Then, a massive gauge boson from hidden $U(1)_D$, dark photon, mediates the interaction between DM and SM particles with tiny coupling induced by kinetic mixing~\cite{Holdom:1985ag}. 

In the case of very 
small coupling between the DM and the SM particles, there is no strong restriction of the coupling for secondary dark particles itself due to annihilation (or decay) with the SM particles at present.
And, if the self-interacting effect of secondary dark particles is also taken into account, it is also expected to behave as a worm DM.
It is allowed to construct a Galactic halo with a certain mixing ratio of primary and secondary DM \cite{Kamada:2021,Kim:2023}. 

For the mass parameter space, the lower mass limit of the DM in the freeze-out process is allowed to be about 5 MeV/$c^{2}$ based on the cosmic microwave background (CMB) observations and the Big Bang nucleosynthesis  (BBN) \cite{DM_CMB:2013,DM_BBN:2019}, and it is interesting because the parameter space on the MeV scale has not been fully investigated in previous direct searches.
%

Since the annihilation process of DM is expected to be concentrated at the center of the Galaxy with high DM density, a more reliable detection method is direction-sensitive detection. Currently, direction-sensitive search experiments are being implemented using low-pressure gas time projection chambers (TPCs) and ultra-high resolution nuclear emulsion \cite{Gas_TPC:2020,NEWSdm:2024}. Directional detection has also been explored for DM searches in neutrino fog, as well as for studying the velocity distribution and density profile of DM \cite{OHare:2021utq,OHare:2014nxd,Kavanagh:2015aqa,Nagao:2017yil,Nagao:2022azp}. 

In particular,  for the lighter mass parameter space around  $O(10-100)$ MeV, lighter targets such as the proton, carbon, fluorine, etc., are more suitable for detection.
In this paper, as an example, we assume a direct detection using nuclear emulsion, NEWSdm experiment \cite{Umemoto:2023hmt}. 
The nuclear emulsion used in the experiment contains relatively light elements, including proton, carbon, and so on, as targets. 
They are more sensitive to light DM and relatively low-energy DM than heavy nuclear targets. Therefore, they are suitable targets for detecting boosted light DM described above.
In Table \ref{Tab:MassFraction}, elements in the nuclear emulsion with mass and their atomic fraction are shown \cite{Asada:2017wvp}. 
For example, by targeting protons, which are the lightest nuclei, DM with a mass of about 1-10 MeV and a kinetic energy of a few MeV, we obtain a recoil energy of a few 10 keV to 100 keV. In such a kinematic parameter space, the recoil of heavy nuclei is difficult to detect because the recoil is less than a few keV. %
Fine-grained nuclear emulsions called the Nano Imaging Tracker (NIT)  have already demonstrated the neutron measurement using the detection of proton tracks in the 100 keV band, which combines quite high $\gamma$- and $\beta$-ray discrimination and direction sensitivity, which cannot be achieved with conventional neutron detectors~\cite{Shiraishi:2023}. Also, this energy threshold with direction sensitivity 
would be improved more below $O(10)$ keV in the future. 

\begin{table}
 \begin{center}
   \caption{Mass and atomic fraction of elements in nuclear emulsion used in NEWSdm}
  \begin{tabular}{ccc}
    \hline
    Element & Mass\% & Atom\% \\ \hline
    Ag & 44.5 & 10.5 \\ 
    Br & 31.8 & 10.1 \\ 
    I & 1.9 & 0.4 \\ 
    C & 10.1 & 21.4 \\ 
    N & 2.7 & 4.9 \\ 
    O & 7.4 & 11.7 \\ 
    p & 1.6 & 41.1 \\
    \hline
  \end{tabular}
 \end{center}
 \label{Tab:MassFraction}
\end{table}

The paper is structured as follows. In Sec. \ref{sec:model}, we present the two-component DM model and its field components. In Sec. \ref{sec:DMscattering}, we introduce the estimation of boosted DM flux and its detection.
The results of the allowed parameter space and the expected event number in the future detection experiment are presented in Sec. \ref{sec:Results}, and we finally conclude in Sec. \ref{sec:Summary}.

\section{A model of two-component DM}
\label{sec:model}

In this section, we show a UV complete two-component DM model with a dark photon mediator that is based on hidden $U(1)_D$ gauge symmetry~\cite{Li:2023fzv}.
In the model, DM candidates are Dirac fermions $\psi$ and $\chi$ with different $U(1)_D$ charges denoted by $Q_\psi$ and $Q_\chi$, respectively.
We also introduce a singlet scalar field $\varphi$ with $U(1)_D$ charge $1$ that develops a vacuum expectation value (VEV) $v_\varphi$ to break the extra gauge symmetry.
Here we impose conditions for charges as $|Q_\chi| \neq |Q_\psi |$, $|2 Q_\chi| \neq 1$, $|2 Q_\psi| \neq 1$, $|Q_\chi| \neq 0$, $| Q_\psi| \neq 0$ and $| Q_\chi \pm Q_\psi | \neq 1$ so that we only have vector portal interaction for DM candidates forbidding Yukawa interactions, and $\chi$ and $\psi$ do not mix. As a result, there remains remnant discrete symmetry $Z_2^\chi \times Z_2^\psi$ in which $\chi$ and $\psi$ are odd under each $Z_2$.
In the scenario, we thus obtain two DM components, $\chi$ and $\psi$, which are both stable due to the discrete symmetry. 

For the $U(1)$ gauge sector, 
the kinetic mixing term is allowed in general
\begin{equation}
\mathcal{L}_{U(1)} = - \frac14 B_{\mu \nu} B^{\mu \nu} - \frac14 X_{\mu \nu} X^{\mu \nu} - \frac{ \epsilon'}{2} X_{\mu \nu} B^{\mu \nu},  
\end{equation}
where $X_{\mu \nu}{=\partial_\mu X_\nu-\partial_\nu X_\mu}$ and $B_{\mu \nu}=\partial_\mu B_\nu-\partial_\nu B_\mu$ are gauge field strength of $U(1)_D$ and $U(1)_Y$, respectively.
After the gauge symmetry breaking, we have massive extra gauge boson $A'$ with mass $m_{A'} \simeq g_D v_\varphi$ assuming tiny kinetic mixing parameter $\epsilon' \ll 1$  where $g_D$ is gauge coupling of $U(1)_D$.
Then $A'$ is identified as a dark photon that has interaction with the SM particles as
\begin{equation}
\mathcal{L}_{A'} = e  \epsilon J^\mu_{\rm EM} A'_\mu,
\end{equation}
where $\epsilon \equiv \epsilon' \cos \theta_W$ with $\theta_W$ being the 
weak mixing angle, $e$ is the electromagnetic coupling constant and $J^\mu_{EM}$ is the electromagnetic current in the SM.

The relevant Lagrangian for new fermions $\chi$ and $\psi$ is 
\begin{equation}
\mathcal{L}_{\rm DM} = \bar \chi (i \slashed{D} - m_\chi) \chi + \bar \psi (i \slashed{D} - m_\psi) \psi, 
\label{eq:fermioninteraction}
\end{equation}
where $D_\mu \chi(\psi) = (\partial_\mu + i Q_{\chi(\psi)} g_D A'_\mu) \chi(\psi)$ is the covariant derivative. 
In the following, we write the DM-$A'$ interaction in eq.(\ref{eq:fermioninteraction}) as 
\begin{equation}
A'_\mu (g_\chi \bar \chi \gamma^\mu \chi + g_\psi  \bar \psi \gamma^\mu \psi),
\end{equation}
where $g_{\chi (\psi)} \equiv Q_{\chi(\psi)} g_D$.

Explaining the relic density of DM in the model, we assume a hierarchy of masses and couplings as $m_\psi > m_\chi \simeq m_{A'}$ and  $g_\psi \ll g_\chi $.
Then, the relic density of DM is determined by annihilation processes 
\begin{align}
&  \psi \bar \psi \to A' \to  \chi \bar \chi, \\
&  \chi \bar \chi \to A' A',
\end{align}
where $ \psi  \bar \psi \to A' A'$ process is subdominant due to adopted coupling hierarchy.
In the early universe, the heavy DM $\psi$ first freezes out due to a smaller annihilation cross section, and $\psi$ becomes the dominant DM component in the current universe.
The relic density of $\chi$ freezes out at a later period, and its relic density is much smaller than that of $\psi$.
The typical value of the cross section to obtain observed relic density is given by $\langle \sigma_{\psi \bar \psi \to \chi \bar \chi} v \rangle \simeq 5 \times 10^{-26} \ {\rm cm}^3/{\rm s}$ that can be obtained by choosing the value of $g_\psi$ relevantly.
For $m_{\psi} \gg m_\chi \simeq m_{A'}$ limit, we can roughly estimate the cross section such that 
\begin{equation}
\langle \sigma_{\psi \bar \psi \to \chi \bar \chi} v \rangle \sim 4.5 \times 10^{-26} \ {\rm cm}^3/{\rm s} \times \left( \frac{g_\chi}{1.0} \right)^2 \left( \frac{g_\psi}{10^{-5}} \right)^2 \left( \frac{20 \ {\rm MeV}}{m_\psi} \right)^2.
\end{equation}
Thus we obtain the observed relic density by choosing $g_\psi \sim 10^{-5}$ when $g_\chi$ is $\mathcal{O}(1)$ and $m_\psi \sim 20$ MeV.

\section{Scattering of boosted DM on nucleus}
\label{sec:DMscattering}
The flux of the DM $\psi$ 
can be estimated as
\begin{equation}
\frac{d\Phi_\mathrm{GC}}{d\Omega d E_\psi}=\frac{1}{4}\frac{r_\mathrm{Sun}}{4\pi}\left(\frac{\rho_\mathrm{local}}{m_\psi}\right)^2 J\langle \sigma_{\psi\bar{\psi}\to \chi\bar{\chi}} v\rangle_{v\to 0} \frac{dN_\chi}{dE_\chi},
\end{equation}
where the distance from the Sun to the earth $r_\mathrm{Sun}=8.33$ kpc, the local DM density $\rho_\mathrm{local}=0.4$ GeV/cm$^3$, and $J$ is halo-shape-dependent parameter \cite{Agashe:2014yua}. 
The flux of $\chi$ from a 10$^\circ$ cone around the Galactic center is
\begin{equation}
\Phi^{10^\circ}_\mathrm{GC}=2.0\times10^{-2}\textrm{cm}^{-2}\textrm{s}^{-1}
C_{\rm pro}\left(\frac{\langle \sigma_{\psi\bar{\psi}\to \chi\bar{\chi}} v\rangle}{5\times 10^{-26} \ \textrm{cm}^3/\textrm{s}}\right)
\left(\frac{60\  \textrm{MeV}}{m_\psi}\right)^2, \label{eq:flux}
\end{equation}
where $C_{\rm pro}$ is a factor depending on DM profile normalized as $C_{\rm pro} =1$ for the Navarro–Frenk–White (NFW) profile \cite{Navarro:1995iw}.
Since the Einasto profile is better favored than the NFW profile as the DM profile of the Galaxy by circular velocity curve, the Einasto profile is assumed hereafter \cite{Jiao:2023aci, Ou:2023adg}.
The Einasto profile adopted in the numerical calculation is
\begin{equation}
\rho_\mathrm{Ein}(r)=\rho_s 
\exp{\left[-\frac{2}{\alpha}\left[\left(\frac{r}{r_s}\right)^\alpha-1\right]\right]},
\end{equation}
where $\alpha=0.11$, $r_s$=35.24~[kpc] and $\rho_s=0.021$~[GeV/cm$^3$] \cite{Cirelli:2010xx}.
For the Einasto profile, we obtain $C_{\rm pro} \simeq 3.76$, and we apply the case in our calculation. 

As discussed in the previous section, the thermally averaged cross section is taken to be $\langle \sigma_{\psi\bar{\psi}\to \chi\bar{\chi}} v\rangle=5\times 10^{-26}$ cm$^3$/s to be consistent with the relic abundance of DM.

With the flux $\Phi^{10^\circ}_\mathrm{GC}$, the number of events $N_{eve}$ detected in the direct detection experiment can be estimated as
\begin{align}
\frac{dN}{d\,\log{E_R}}&=E_R\frac{dN}{dE_R}=\Delta T N_\mathrm{target}\Phi^{10^\circ}_\mathrm{GC} E_R \frac{d \sigma_{\chi N\to \chi N}}{dE_R} \nonumber,\\
N&= \Delta T N_\mathrm{target}\Phi^{10^\circ}_\mathrm{GC} \int d E_R \frac{d \sigma_{\chi N\to \chi N}}{dE_R}
\label{eq:dNdER}
\end{align}
where $\Delta T$ is 
the exposure time, 
$N_\mathrm{target}$ is the number of targets in the detector, and $E_R$ is the recoil energy of target nucleus.
The energy differential cross section of $\chi$-nucleus scattering is
\begin{align}
\int_{E_R^{\rm min}}^{E_R^{\rm max}} d E_R \frac{d \sigma_{\chi N\to \chi N}}{dE_R}&= \int_{E_R^{\rm min}}^{E_R^{\rm max}} d E_R \left| \frac{d\Omega}{dE_R} \right| \frac{d\sigma_{\chi N\to \chi N}}{d\Omega}, 
\label{eq:cx-trans}
\end{align}
where $E_R^{\rm min(max)}$ is the maximal(minimal) recoil energy.
The angular differential cross section for $\chi ({\bf p}) N ({\bf 0}) \to \chi ({\bf p}') N({\bf q})$ process is represented with the electric and magnetic Sachs form factors $G_E$ and $G_M$ as~\cite{Agashe:2014yua}
\begin{align}
\frac{d\sigma_{\chi N\to \chi N}}{d\Omega}=
&\frac{1}{(4\pi)^2}\frac{(\epsilon e)^2g^{'2}}{(q^2-m_{A'}^2)^2}\frac{p'/p}{1+(E_\chi-p E'_\chi \cos{\theta}/p')/m_N} \nonumber \\
&\times \left[G_E^2\frac{4E_\chi E'_\chi+q^2}{1-q^2/(4 m_N^2)}+
G_M^2\left((4E_\chi E'_\chi+q^2)\left(1-\frac{1}{1-q^2/(4m_N^2)}\right)
+\frac{q^4}{2m_N^2}+\frac{q^2m_\chi^2}{m_N^2}\right)\right] \label{Eq:chiNtochiN}\\
&G_E(q^2)=\frac{G_M(q^2)}{2.79}=\frac{1}{(1+q^2/(0.71 \mathrm{GeV}^2))^2}\nonumber
\end{align}
where $m_N$ is the target nucleus mass and $\theta$ is the scattering angle.

In direct detection, events with recoil energy above the detector's energy threshold can be detected.
Note that in this paper, we are going to study the case where $\psi$ is non-relativistic DM 
in Sec\ref{sec:Results}. 
{Thus, from the energy-momentum conservation, the relation between scattering angle and the recoil energy is represented as
\begin{align}
\cos \theta = \frac{m_\psi^2 - m_\chi^2 - (m_\psi + m_N) E_R}{\sqrt{m_\psi^2 - m_\chi^2} \sqrt{(m_\psi - E_R)^2 - m_\chi^2}}.
\label{eq:cx-transform}
\end{align}
The factor of $|d\Omega/dE_R|$ in Eq.~\eqref{eq:cx-trans} can be obtained from the relation.
In the limit of 
$m_\chi \ll m_\psi$ and $m_\psi \ll m_N$ it can be approximated as 
\begin{equation}
\cos \theta \simeq 1 - \frac{m_N E_R}{m_\psi^2},
\end{equation}
and $|d\Omega/dE_R| \simeq 2 \pi m_N/m^2_\psi$.
Furthermore, maximal recoil energy can be estimated, in the case of $m_\psi \ll m_N$, such that 
\begin{equation}
\label{eq:ERmax}
E_R^{\rm max} \simeq 2\frac{m_\psi^2 - m_\chi^2}{m_\psi+m_N} = 2\frac{m_\psi + m_\chi}{m_N + m_\psi} \delta m,
\end{equation}
where $\delta m \equiv m_\psi - m_\chi$.
}

 \begin{figure}[tb]
 \begin{center}
\includegraphics[width=8cm]{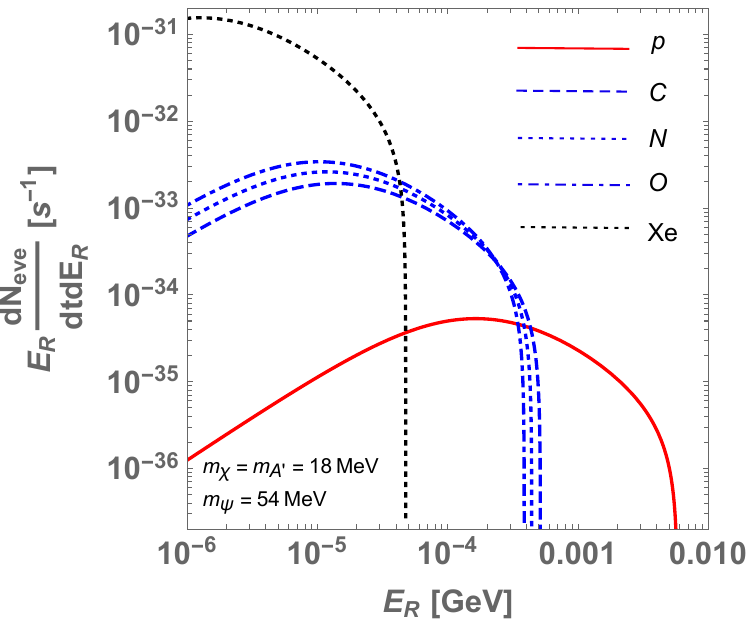} \
\includegraphics[width=8cm]{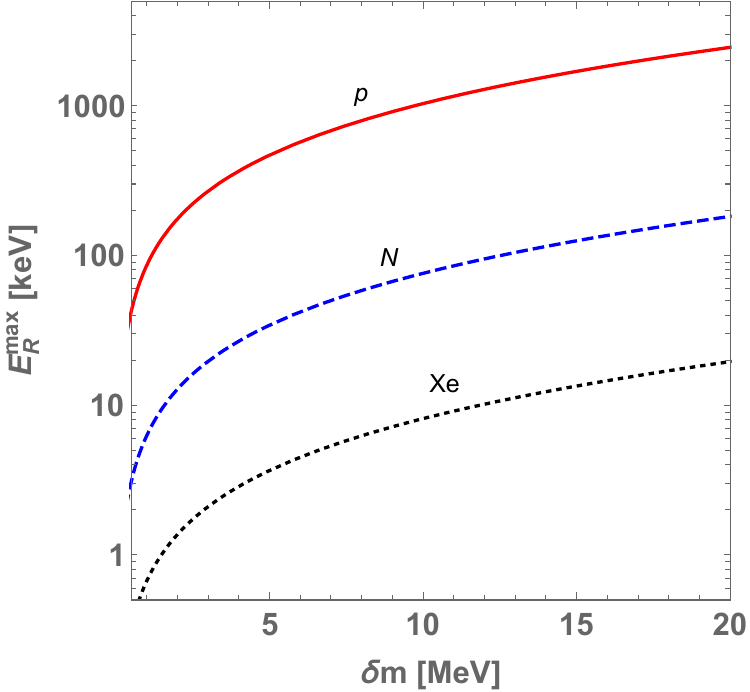}
\caption{Left: The function associated to the event rate $E_R \frac{dN_{eve}}{d E_R d t}$ as functions of $E_R$ for each nucleus where we adopt $m_\chi = m_{A'} = 18$ MeV, $m_\psi = 3 m_\chi$, $g_X = 1$ and $e \epsilon = 2 \times 10^{-4}$ as a reference point. Right: Maximal recoil energy for each nucleus as a function of $\delta m$ where we adopt $m_\chi = m_{A'} = 20$ MeV.} 
\label{fig:dist}
\end{center}
\end{figure}

\begin{figure}[tb]
 \begin{center}
\includegraphics[width=8cm]{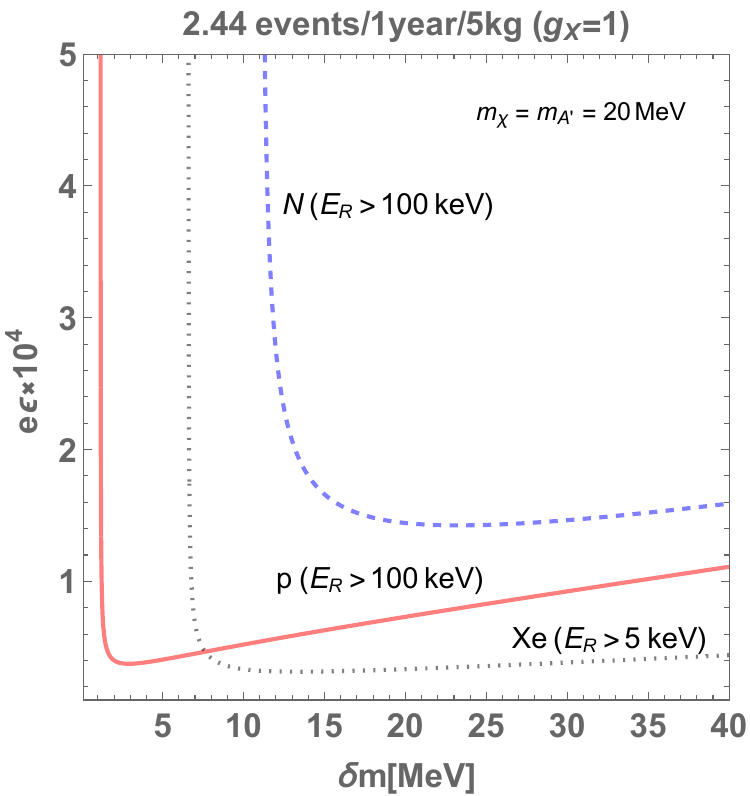} \
\includegraphics[width=8cm]{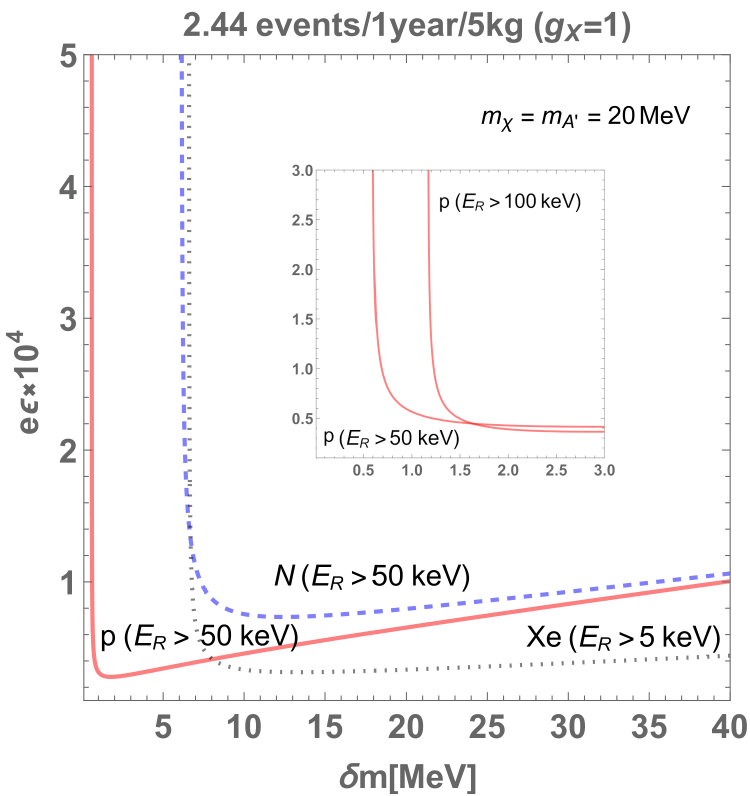} 
\caption{Contours giving $N_{ev} =2.3$ on $\{\delta m, e \epsilon\}$ plane where $m_\chi = m_{A'} = 20$ MeV. The threshold energy is 100 keV and 50 keV for p and N in the left and right figures, while that of Xe is taken to be 5~keV. In the right plot, we also show  the behavior of the proton case with different threshold energy in small $\delta m$ region.}
\label{fig:contour}
\end{center}
\end{figure}

For illustration, we show $E_R \frac{dN_{eve}}{d E_R d t}$ for each nucleus as functions of $E_R$ in the left plot of Fig.~\ref{fig:dist} for the case that $\chi$ is relativistic. 
We find that the distribution for p shows a peak around 200 keV while the distributions for other nuclei have peaks around 10~keV due to heavier masses than p. 
Note that even at $E_R>400$ keV, where C, N, and O lose detectability, only p is still detectable.
In addition, maximal recoil energy for each nucleus is shown in the right plot of Fig.~\ref{fig:dist} as a function of $\delta m$ from Eq.~\eqref{eq:ERmax} adopting $m_\chi = m_\psi = 20$ MeV; we omit lines for C and O since they are almost the same as that of $N$.
In other words, protons are a preferred target because they are likely to be detectable even at a relatively high detector energy threshold.
Furthermore, in Fig.~\ref{fig:contour} we show contours giving number of event 
$N_{eve} =2.44$, which corresponds to background-free 90\% C.L. exclusion limit \cite{Feldman:1997qc}, on $\{\delta m, e \epsilon\}$ plane where $m_\chi = m_{A'} = 20$ MeV is chosen. 
We find that only the proton target has sensitivity when the mass difference of DMs is sufficiently as small as $\simeq $6~MeV.

\section{Results}
\label{sec:Results}

 \begin{figure}[tb]
 \begin{center}
\includegraphics[width=8cm]{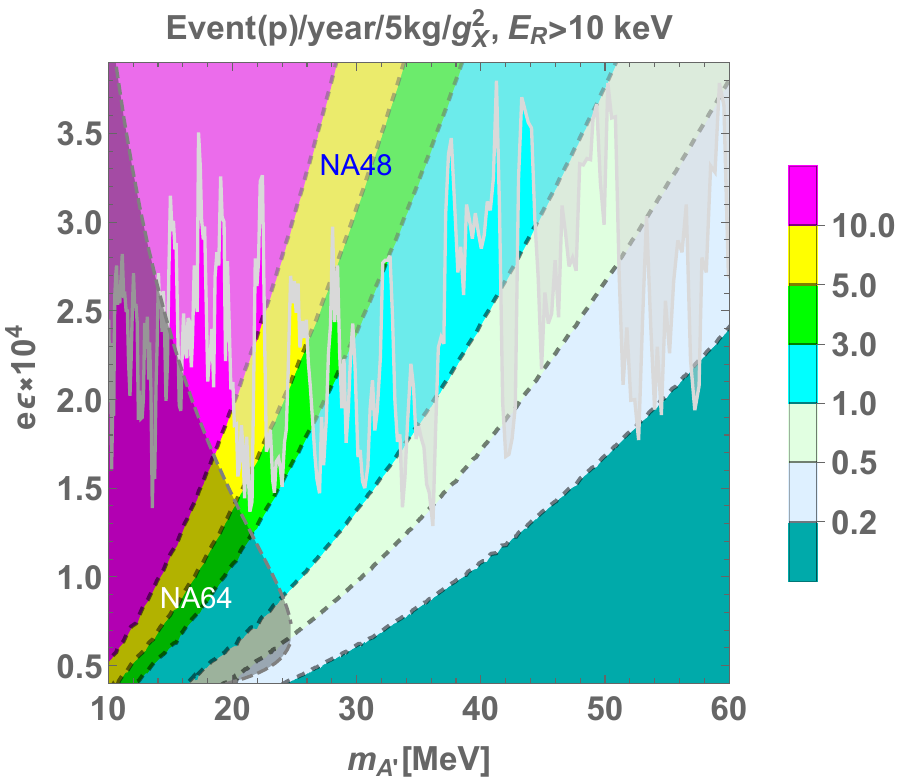} \ 
\includegraphics[width=8cm]{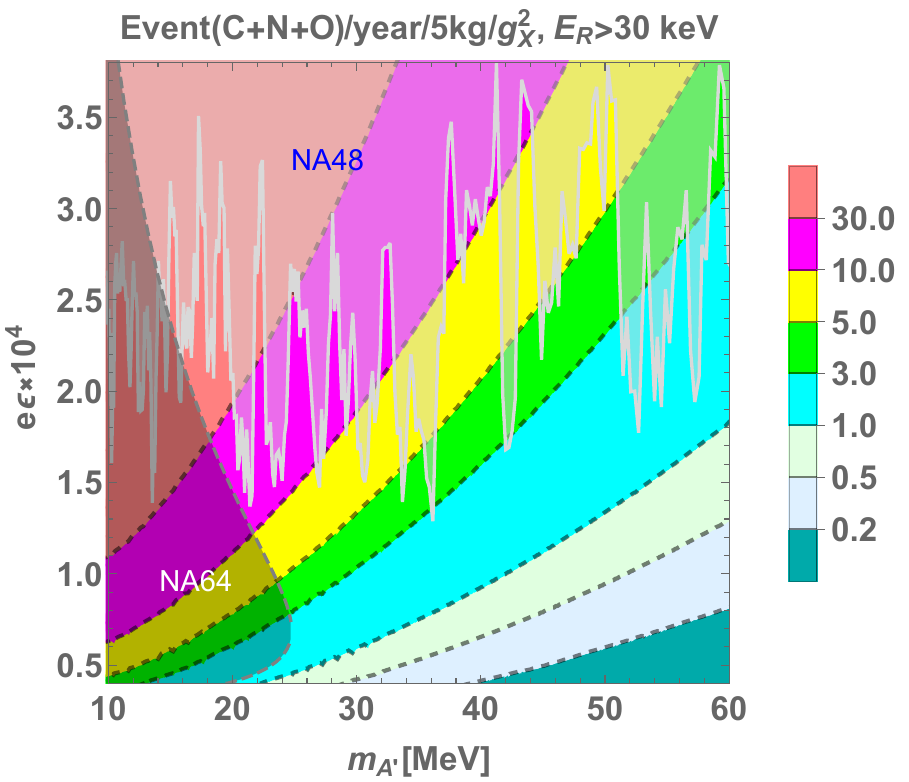}  
\caption{Estimated number of events per year per 5 kg for the ideal energy threshold assuming $g_\chi=1$. The left (right) plot corresponds to the case of the target of detector p (C, N, and O).
Areas shaded pale and shaded dark are excluded by NA48 and NA64, respectively.}
\label{fig:Events1}
\end{center}
\end{figure}

 \begin{figure}[tb]
 \begin{center}
\includegraphics[width=8cm]{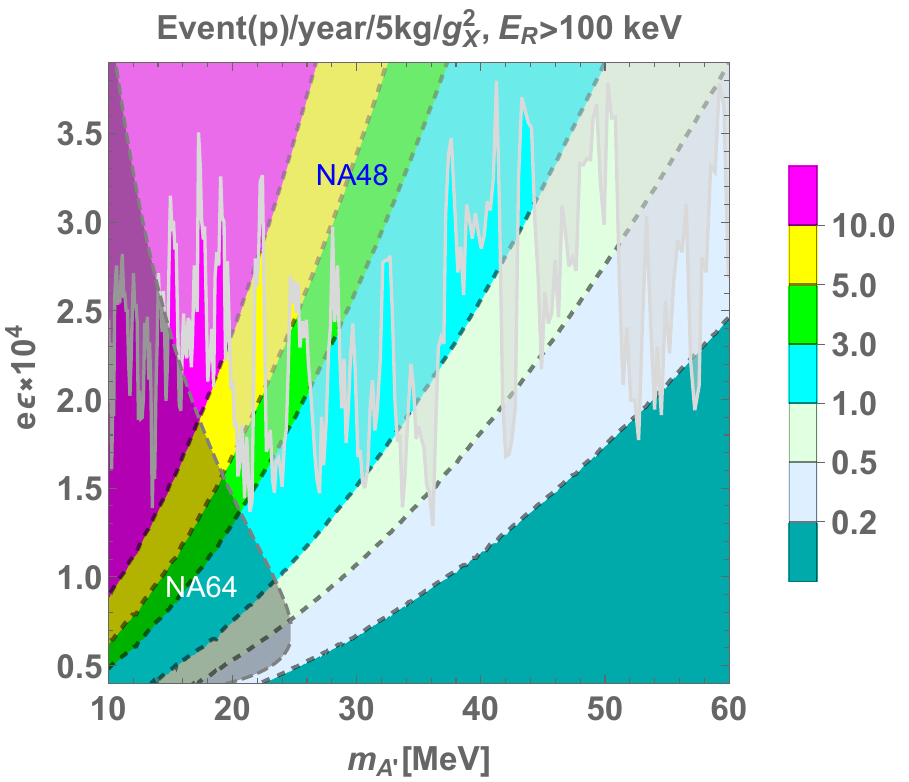} \ 
\includegraphics[width=8cm]{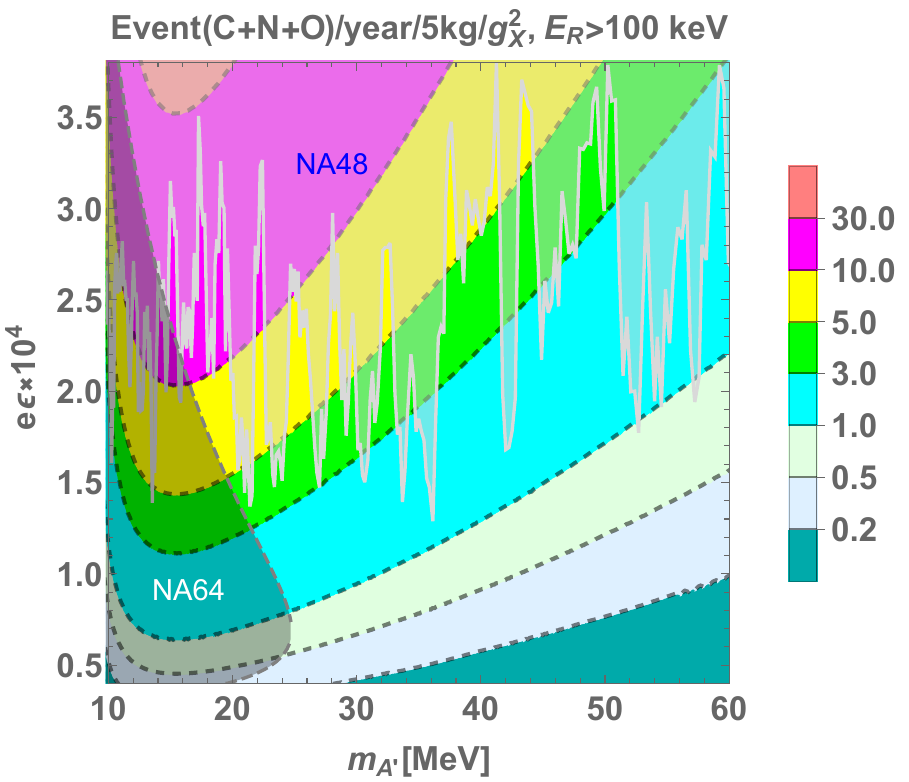}  
\caption{Estimated number of events. Legend is the same as Fig.~\ref{fig:Events1} except the energy thresholds are assumed to be practical.}
\label{fig:Events2}
\end{center}
\end{figure}

In this section, we show the expected number of events for scattering between boosted DM and light nuclei \{p, C, N, O\} that is estimated by formalism in the previous section.
In our calculation, we fix the mass relation between $\chi$ and $A'$ as follows
\begin{equation}
m_\chi = m_{A'}.
\end{equation}
Here we consider degenerate mass for $\chi$ and $A'$ to suppress $A'$ radiation from $\chi$ after production via annihilation of $\bar \psi \psi $ pair to avoid indirect detection constraint~\cite{Li:2023fzv}.
 From Eq.~(\ref{Eq:chiNtochiN}), it is clear that the scattering cross section $\sigma_{\chi N\to \chi N}$ does not depend much on $m_{A'}$. Therefore, even if $m_{A'}$ is changed, the results of this study are almost unaffected.
For $\psi$ mass, we consider two cases 
\begin{align}
& \text{Hierarchical case:} \quad  m_\psi = 3 m_\chi, \\
& \text{Quasi-degenerate case:} \quad m_\psi = m_\chi + \delta m,
\end{align}
where we choose $\delta m = 5$ MeV.
For the hierarchical case, we obtain sufficiently boosted $\chi$ from $\psi \bar \psi$ annihilation, and $\chi$ can be detected by any target nuclei. 
On the other hand, in the quasi-degenerate case, only the proton target has sensitivity, as indicated by Fig.~\ref{fig:contour}.
We then focus on mass range suitable to discuss scattering of DM and light nuclei that is 
\begin{equation}
m_\chi \in [10, 60] \, {\rm[MeV]}.
\end{equation}
In the mass range, we need to take into account the constraints of dark photon search.
The relevant bounds on $\{m_{A'}, e \epsilon \}$ plane are obtained from NA64~\cite{NA64:2018lsq,NA64:2019auh} and NA48 \cite{NA482:2015wmo} experiments.
In addition, we consider two schemes for a threshold of nuclei recoil energy to obtain the detectable range of 
the nuclear recoil
\begin{align}
& \text{Scheme (i): } \quad \, {\rm p :} \ E_R > 10 \ {\rm keV}, \quad {\rm C,N,O:} \ E_R > 30 \ {\rm keV}, \\
& \text{Scheme (ii): } \quad {\rm p, C, N, O:} \ E_R > 100 \ {\rm keV}.
\end{align} 
The scheme (i) is chosen to obtain a track length on nuclear emulsion larger than 100 nm, which corresponds to the ideal energy threshold to aim for going forward, 
while scheme (ii) corresponds to a case already at a practical level. 

\subsection*{Hierarchical case}

In Figs.~\ref{fig:Events1} and \ref{fig:Events2}, we show the expected number of events for light nuclei per year and 5 kg target with $g_X = 1$ on $\{m_{A'}, e \epsilon \}$ plane, in the schemes (i) and (ii), respectively. 
The color gradient indicates the number of events, as shown in the figure. 
The gray and light gray regions are excluded by NA64 and NA48 dark photon search results, respectively.
We find that the detectable number of events can be realized in some allowed parameter region.
Also, it is found that the number of events decreases for relatively heavier nuclei in the scheme (ii) due to the $E_R$ distribution shown in Fig.~\ref{fig:dist}. 

In addition, we show the expected number of events per year for benchmark points (BPs) considering 5 kg and 50 kg nuclear emulsion detectors in the schemes (i) and (ii), respectively,  
where we expect a larger amount of nuclear emulsion in the scheme (ii) as the $E_R$ threshold is practical.
The nuclear emulsion contains light nuclei p, O, N, and C with mass fractions of 1.6 \%, 7.4 \%, 2.7 \%, and 10.1 \%, respectively.
In the benchmark scenarios, we expect 
several numbers of events per year.
Note also that light nuclei \{p, C\} are more promising for scheme (ii) since they tend to provide larger $E_R$. 

\begin{table}[t]
\begin{tabular}{c||c c c c c c c}\hline  
\multicolumn{7}{c}{Scheme (i), 5 (50) kg nuclear emulsion} \\ \hline
&  ~$\{m_\chi/{\rm MeV}, \ \epsilon, \ g_\chi \}$~ & ~$N_{\rm p}$~ & ~$N_{\rm C}$~ & ~$N_{\rm N}$~ & ~$N_{\rm O}$~ & ~$N_{\rm Sum}$~  \\\hline
\hspace{2mm}BP1\hspace{2mm}  & $\{ 24, 1.4\times 10^{-4}, 2.5 \} $ & $0.25 \ (2.5)$  & $1.5 \ (15)$ & $0.57 \ (5.7)$ & $2.2 \ (22)$  & $4.4 \ (44)$     \\
\hspace{2mm}BP2\hspace{2mm}   & $\{18, 2.0\times 10^{-4}, 2.0 \} $ & $1.0 \ (10)$  & $3.9 \ (39)$ & $1.5 \ (15)$ & $5.5 \ (55)$ & $12 \ (1.2 \times 10^2)$ \\\hline
\multicolumn{7}{c}{Scheme (ii), 50 (100) kg nuclear emulsion} \\ \hline
&  ~$\{m_\chi/{\rm MeV}, \ \epsilon, \ g_\chi \}$~ & ~$N_{\rm p}$~ & ~$N_{\rm C}$~ & ~$N_{\rm N}$~ & ~$N_{\rm O}$~ & ~$N_{\rm Sum}$~  \\\hline
\hspace{2mm}BP1\hspace{2mm}  & $\{ 24, 1.4\times 10^{-4}, 2.5 \} $ & $1.9 \ (3.8)$  & $5.1 \ (10)$ & $1.8 \ (3.9)$ & $0.65 \ (1.3)$  & $9.6 \ (19)$     \\
\hspace{2mm}BP2\hspace{2mm}   & $\{18, 2.0\times 10^{-4}, 2.0 \} $ & $6.3 \ (13)$  & $9.8 \ (20)$ & $3.3 \ (6.6)$ & $1.1 \ (2.2)$ & $21 \ (42)$ \\\hline
\end{tabular}
\caption{Expected number of events for hierarchical case in the benchmark points per year for scattering between DM and \{p, C, N, O\} with 5 (50) kg and 50 (100) kg nuclear emulsion in the scheme (i) and (ii) respectively.
We adopt mass fraction in nuclear emulsion as 1.6 \%, 7.4 \%, 2.7 \%, and 10.1 \% for p, O, N, and C, respectively. In the right column, the total number of event $N_{\rm Sum} = N_{\rm p} + N_{\rm C}+ N_{\rm N}+ N_{\rm O}$ is shown.} 
\label{tab:2}
\end{table}

Here, we would like to emphasize that measurements of DM-nucleon interaction by nuclear emulsion 
test the combination of DM-$A'$ and nucleon-$A'$ interactions while constraints on dark photon interactions by NA48 and NA64 involves $A'$-quark and $A'$-electron interactions
as these experiments detect electrons from $A' \to e^+ e^-$.
Thus, the measurements by nuclear emulsion provide an independent test of $A'$ interactions associated with nucleon and DM. 
In fact, it would be possible to relax constraints from NA48 and NA64 if we consider some exotic $A'$ interaction whose behavior changes from dark photons.
It is therefore important to test $A'$ interactions by various experiments measuring different processes.

 \begin{figure}[t!]
 \begin{center}
\includegraphics[width=8cm]{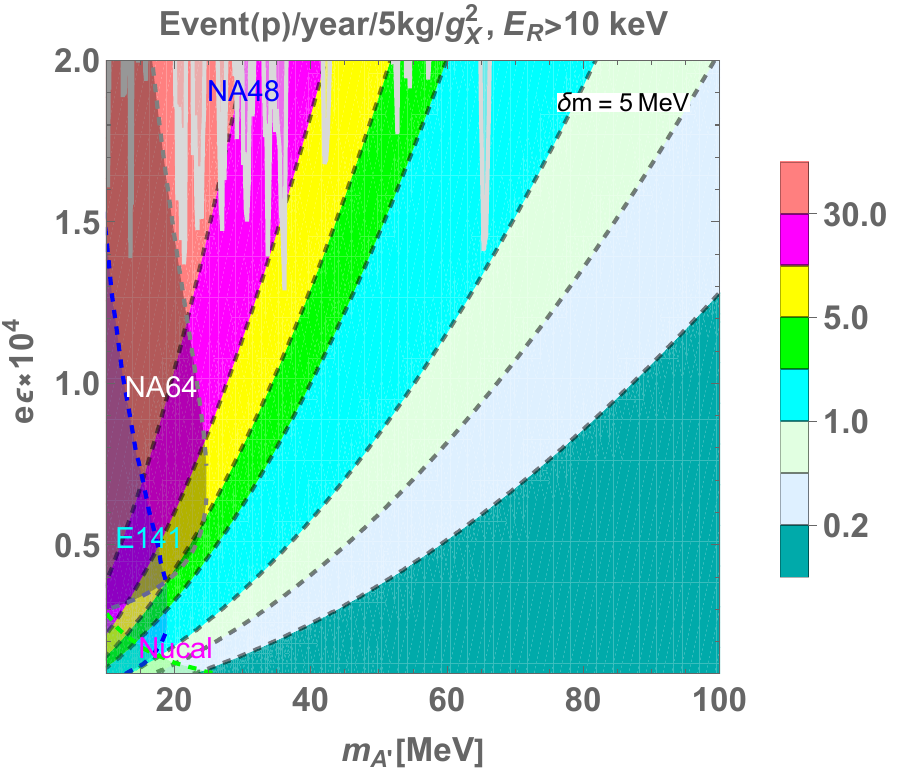} \ 
\includegraphics[width=8cm]{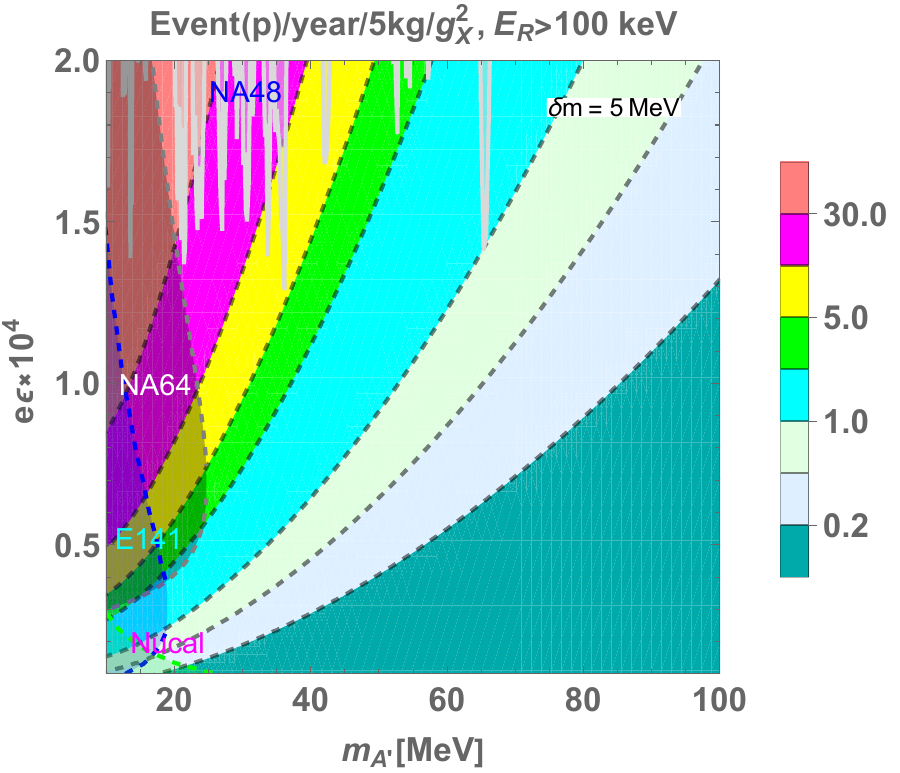} 
\caption{Estimated number of events for the quasi-degenerate case. Legend is same as Fig.~\ref{fig:Events2}. The target of the detector is assumed to be protons. The left (right) plot corresponds to the ideal (practical) energy threshold of the detector.}
\label{fig:Events3}
\end{center}
\end{figure}

\subsection*{Quasi-degenerate case}

In the left and right figures in Fig.~\ref{fig:Events3}, we show the expected number of events in the quasi-degenerate case for proton target on $\{m_{A'}, e \epsilon \}$ plane, in the schemes (i) and (ii), respectively.
The color gradient indicates the number of events, as shown in the figure. 
In addition to NA64 and NA48 constraints in Figs.~\ref{fig:Events1} and \ref{fig:Events2}, we added constraints from E141 \cite{Riordan:1987aw} and Nucal \cite{Blumlein:1991xh,Blumlein:2013cua,Tsai:2019buq} experiments since we extended the region of $e \epsilon$.
We obtain a larger number of events than the hierarchical case since $m_\psi$ is smaller and $\chi$ flux is enhanced as understood by Eq.~(\eqref{eq:flux}).
The number of events in scheme (ii) is slightly smaller than in scheme (i) due to the larger energy threshold.

Furthermore, we show the expected number of events in quasi-degenerate case per year for the same BPs and schemes in Table~\ref{tab:3}. 
In this case we only consider p target with 1.6 \% mass fraction in the nuclear emulsion.
We find a much larger number of events than table~\ref{tab:2} for the same BPs due to the enhancement of $\chi$ flux as discussed above.
Thus, smaller $g_X$ and/or $\epsilon$ regions can be tested in the case.

\begin{table}[t]
\begin{tabular}{c||c c }\hline  
&  ~Scheme (i), 5 (50) kg~ & ~Scheme (ii), 50 (100) kg~  \\\hline
\hspace{2mm}BP1\hspace{2mm}  & $32 \ (3.2 \times 10^2)$ & $1.9 \times 10^2 \ (3.8 \times 10^2)$   \\
\hspace{2mm}BP2\hspace{2mm}  & $1.0 \times 10^2 \ (1.0 \times 10^3)$ & $4.7 \times 10^2 \ (9.4 \times 10^2)$ \\\hline
\end{tabular}
\caption{Expected number of events for quasi-degenerate case in the same BPs as Table~\ref{tab:2} per year for scattering between DM and p with 5 (50) kg and 50 (100) kg nuclear emulsion in the scheme (i) and (ii) respectively}\label{tab:3}
\end{table}

\subsection*{Output from directional dark matter search}
\begin{figure}[!h]
 \begin{center}
\includegraphics[width=12cm]{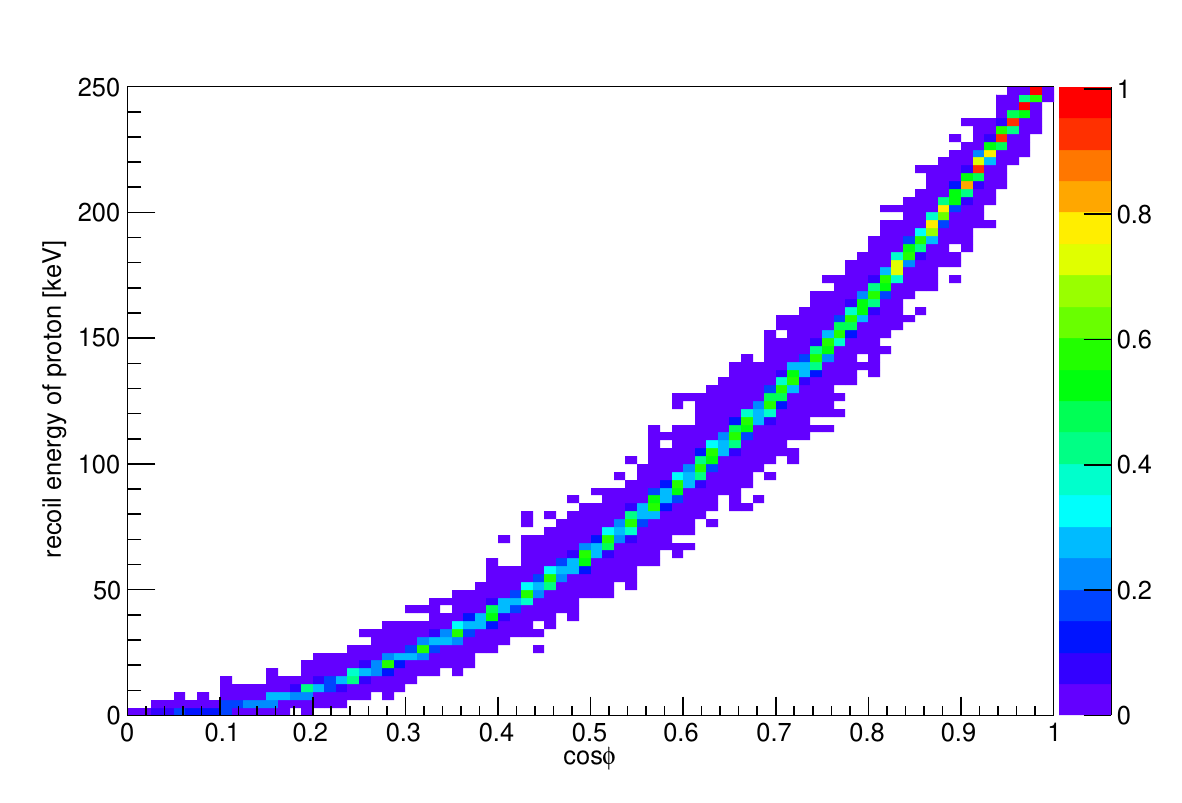}
\caption{
Distribution of recoil energy and cosine of the scattering angle $\phi$ relative to the Galactic Center for proton recoils from $\chi$-p elastic scattering. The plot assumes an Einasto profile, $m_\chi= 10$ MeV, and $\delta m = 5$ MeV for the quasi-degenerate case.}
\label{fig:directional}
\end{center}
\end{figure}

To conclude, we will examine the boosted dark matter signals in detectors with directional sensitivity.
For elastic scattering involving target nuclei, assuming dark matter–baryon interaction, the distribution of energy and $\cos{\phi}$ is expected to follow eq.(\ref{eq:cx-transform}) and show the pattern presented in Fig.\ref{fig:directional} for the quasi-degenerate case with $\delta m = 5$ MeV. Here $\phi$ represents the scattering angle relative to the axis connecting the Galactic Center and Earth.
This distinctive signal arises due to its directional sensitivity. The signal originating from the Galactic Center, characterized by a unique incident kinematic energy spectrum, can be distinguished from isotropic background events, which should exhibit a uniform distribution in $\cos{\phi}$. 
For example, in the NEWSdm experiment, this measurement can be performed by keeping the detector aligned with the Galactic Center using an equatorial telescope. 
A similar analysis has been demonstrated using monochromatic neutrons as a point-like source \cite{A4}.

\section{Summary and discussion}
\label{sec:Summary}
Models with two DM components in which the light DM is boosted by heavy DM annihilation are interesting 
because the light DM is detectable even if its mass is below the GeV scale.
Since such annihilation is likely to occur in the Galactic center, where DM density is high, boosted DM events from the Galactic center can be expected. Therefore, direct detection with directional sensitivity is suitable for the verification of the model.

In the paper, we assume a two-component DM model with $U(1)_D$ dark gauge symmetry as a specific example and investigate the directionally sensitive direct detection for MeV-scale DM boosted by annihilation. 
We mainly focus on a nuclear emulsion detector with multiple target atoms included.
For light DM and dark photons with $O(10)$ MeV masses, typically, 
$O(1-10)$ events can be expected with a 5 kg-yr exposure. 
The sensitivity to the boosted DM is also examined by different targets used in the NEWSdm detector. Especially if the two DM masses are degenerate, proton is a better suited target to verify the boosted DM than xenon, the target employed in existing detectors. Note that xenon may have higher sensitivity in the non-degenerate case; however, it does not have directional sensitivity in any case.

Let us discuss in detail the technical aspects of the validation of the boosted DM coming from the Galactic center with NEWSdm.
In particular, proton recoil tracks around 100 keV have been demonstrated to allow 3-dimensional (3D) reconstruction~\cite{Shiraishi:2023}. Also, lower energy tracks, such as proton, C, N, and O recoils around several tens of keV, have shown successful 2-dimensional (2D) tracking, and it is anticipated that these tracks can be reconstructed in 3D using super-resolution microscopy because this limitation is due to the microscope resolution, not the detector itself~\cite{A1,A2}. For 3D reconstruction, a unique distribution between energy and $\cos{\theta}$ is expected as described in eq.(\ref{eq:cx-transform}). For the 2D reconstruction case, it still has the advantage of identifying the signal with a distinct angular distribution in the azimuth angle~\cite{A3}. Such direction-sensitive detection has a powerful advantage for identifying the signal coming from a specific direction, such as the Galactic center, for this scenario. 

Similar measurements can also be performed using gaseous detectors~\cite{A4}. Current gaseous detectors used in directional DM searches employ gases such as CF$_4$ and SF$_6$, among others. The mass of these targets is comparable to the 
C, N, and O targets in this study, so similar sensitivities are expected.

\section*{Acknowledgement}
The work was in part supported by JSPS KAKENHI Grant Numbers No.~21K03562, No.~21K03583, No.~24K07061, No.~24K07032 and No.~24H02244  (KIN), No.~19H05806, No.~23KK0058, No.~24H02236, No.~24H02241, No.~24A205 (T.~Naka), and by the Fundamental Research Funds for the Central Universities (T.~Nomura). 
We would like to thank the Workshop on Inelastic Nuclear Scattering for Dark Matter Detection (held at Kurashiki), where some of the work has been carried out.
%

\end{document}